\documentclass[12pt]{iopart}

\usepackage{graphicx}
\usepackage{textcomp}
\usepackage{color}

\newcommand{\ket}[1]{|{#1}\rangle}

\newcommand{\braket}[2]{\langle {#1} | {#2} \rangle}

\def\opex{ Opt.\ Express }

\def\apl{ Appl.\ Phys.\ Lett.\ }

\def\prl{ Phys.\ Rev.\ Lett.\ }

\def\sci{ Science }
\def\sciadv{ Sci.\ Adv.\ }
\def\scirep{ Sci.\ Rep.\ }
\def\njp{ New.\ J.\ Phys.\ }
\def\natpho{ Nat.\ Photon.\ }
\def\natphy{ Nat.\ Phys.\ }

\begin{document}

\title[Generation of entangled photons using an arrayed waveguide grating]{Generation of entangled photons using an arrayed waveguide grating}

\author{Nobuyuki Matsuda$^{1,2,*}$, Hidetaka Nishi$^{1,3}$, Peter Karkus$^{2}$, Tai Tsuchizawa$^{1,3}$, Koji Yamada$^{1,3}$, William John Munro$^{2}$, Kaoru Shimizu$^{2}$, Hiroki Takesue$^{2}$}

\address{$^1$NTT Nanophotonics Center, NTT Corporation, 3-1 Morinosato-Wakamiya, Atsugi, Kanagawa 243-0198, Japan}
\address{$^2$NTT Basic Research Laboratories, NTT Corporation, 3-1 Morinosato-Wakamiya, Atsugi, Kanagawa 243-0198, Japan}
\address{$^3$NTT Device Technology Laboratories, NTT Corporation, 3-1 Morinosato-Wakamiya, Atsugi, Kanagawa 243-0198, Japan}
\ead{$^*$m.nobuyuki@lab.ntt.co.jp}
\vspace{10pt}

\begin{abstract}
We propose an on-chip source of entangled photon pairs that uses an arrayed-waveguide grating (AWG) with multiple nonlinear input waveguides as correlated photon pair sources. The AWG wavelength-demultiplexes photon pairs created in input waveguides and simultaneously produces a high-dimensional entangled state encoded in the optical path. We implemented the device with a monolithic silicon-silica waveguide integration platform and demonstrated the entanglement of two dimensions in a proof-of-principle experiment.
\end{abstract}

\section{Introduction}

Generation of quantum-entangled photon pairs lies at the heart of photonic quantum information science and technologies \cite{nielsen00}. Recently, entangled states of high-dimensional quantum systems, \textit{i.e.}, qudits, are drawing much attention, since such systems enable us to improve the robustness of quantum key distribution \cite{cerf02}, improve the generation rate of quantum random numbers \cite{xu16}, simplify quantum logic \cite{lanyon09}, and test the foundations of quantum mechanics such as quantum contextuality \cite{hu16}. Motivated by these prospects, the generation of high-dimensional quantum entanglement of photons has been extensively investigated in various physical degrees of freedom of light including orbital-angular momentum \cite{dada11, krenn14, zhang16}, time-bin \cite{riedmatten02, ikuta16}, frequency \cite{xie15, kues17, imany17}, and spatial (path) information \cite{schaeff12, schaeff15}.

Recently, integrated photonic waveguides have proved to be versatile tools for conducting photonic quantum information experiments \cite{politi08, peruzzo10, spring13, spagnolo14, carolan15}. In such a platform, by encoding the quantum states in the optical paths of photons, we can fully exploit the benefits of integrated photonic circuits. For example, an arbitrary unitary transformation of a path-encoded $N$ dimensional quantum system can be realized in a reconfigurable way with $N$ dimensional universal linear optics circuit \cite{carolan15, reck94, harris16}. Furthermore, it was recently shown that such path-encoded qudits are compatible with transmissions over multi-core optical fibers \cite{ding16, canas16}. By virture of these characteristics, the optical path is an attractive physical degree of freedom for generation of high-dimensional and entangled states of photons. Schaeff \textit{et al.} \cite{schaeff12, schaeff15} recently proposed a scheme to generate a high-dimensional path-entangled state using on-chip photon pair sources and fiber optics. However, to take advantage of integrated photonic circuits, it would be ideal to have an entanglement source in a on-chip platform that has the capability to be monolithically integrated with other quantum circuits such as universal linear optics circuits \cite{carolan15}.

In this paper, we describe a compact, on-chip scheme for generating path-encoded high-dimensional entanglement using $N$ multiple photon pair sources and a wavelength demultiplexer using an arrayed waveguide grating (AWG). An AWG is an on-chip device that is widely used as a wavelength (de)multiplexer in current optical-fiber communication networks\textcolor{black}{, as well as frequency multiplexers for entanglement generation \cite{matsuda12} like other (dense) \textcolor{black}{wavelength division multiplexing} (WDM) filters \cite{kaiser12, autebert16}}. The AWG also has the capability to be monolithically integrated with waveguide interferometers \cite{suzuki98}, which are commonly used in on-chip photonic quantum circuits. We use the AWG to wavelength-demultiplex correlated photons generated in the photon pair sources and simultaneously multiplex the photons into a high-dimensional path-entangled quantum state. We fabricated the essential part of the device by using a silicon-silica monolithic photonic integration platform \cite{nishi10} and performed a proof-of-principle experiment for $N$ = 2.

\section{Device design}

Figure 1 is a conceptual schematic of a source for generating $N$-dimensional path-entangled photon pairs. The device consists of a $1 \times N$ splitter, $N$ nonlinear waveguides (photon pair sources), and an AWG. A pump pulse with a wavelength of $\lambda_\mathrm{p}$ is divided equally into $N$ nonlinear waveguides with the $1 \times N$ splitter. In each nonlinear waveguide, a correlated pair of photons having \textcolor{black}{non-degenerate} wavelengths of $\lambda_\mathrm{s}$ (signal) and $\lambda_\mathrm{i}$ (idler) are generated through a nonlinear process such as spontaneous parametric down conversion or spontaneous four wave mixing (SFWM). The nonlinear waveguides are used as, or are directly connected to, the input waveguides of the AWG. The AWG consists of two focusing slab regions (the first and second slabs in the figure) and a phased array of multiple waveguides (waveguide array). Because of the \textcolor{black}{WDM} capability of the AWG, the focal spots of signal and idler modes are spatially separated on the output end of the second slab (arc o-o'). Furthermore, the correlated photons created in each of the input waveguides are focused onto different output positions. The output waveguides A$_j$ and B$_j$ are connected at the end face of the second slab in order to collect the signal and idler photons generated in the $j$-th photon pair source. At the end of the device, we obtain an entangled state of the path-encoded $N$-level system as follows:

\begin{equation}
	\ket{\Psi} = \frac{1}{\sqrt{N}} \sum_{j = 1}^N \mathrm{e}^{-i \phi_j} \hat{a}_{j}^\dagger \hat{b}_{j}^\dagger \ket{0},
\end{equation}
\noindent
where $\phi_j$ is the relative phase that can be tuned by the phase shifters, and $\hat{a}_{j}^\dagger$ and $\hat{b}_{j}^\dagger$ are the creation operators for photons in waveguide modes A$_j$ and B$_j$, respectively. We omit the vacuum and higher order terms for simplicity.

\begin{figure}[tb]
\centerline{\includegraphics[width=14cm]{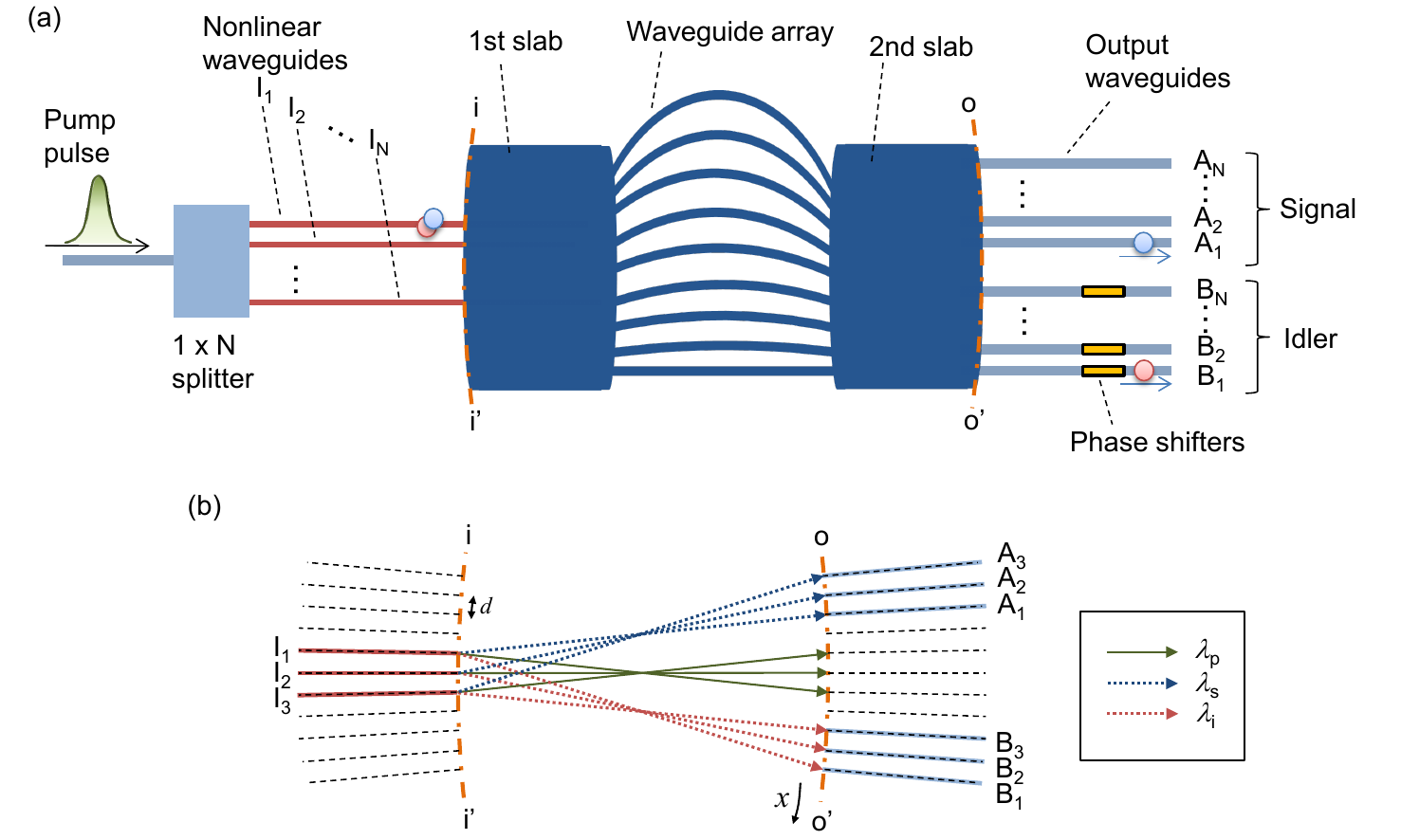}}
\caption{(a) Concept of arrayed waveguide grating source of path-entangled photons. (b) Relationship between the focal positions of target frequency components on the input facet of the first slab (i--i') and output facet of the second slab (o--o') for $N = 3$. The dashed lines show the normals to each facet at a constant spacing $d$.}
\end{figure}

Figure 1(b) shows an example of the i/o waveguide arrangement of the AWG for $N = 3$. Here, we consider the SFWM for the photon pair generation process in the nonlinear waveguides because SFWM efficiently occurs in integrated waveguides such as silicon waveguides \cite{matsuda12, sharping06, takesue07, silverstone14}, which can be integrated with an AWG on a chip \cite{matsuda14, matsuda16}. The black dashed lines coming from the arcs i-i' and o-o' show the grids with equal pitch of $d$, \textit{i.e.}, equal wavelength spacing $\Delta \lambda = \frac{n_{\mathrm{s}} \lambda_0 d^2}{n_\mathrm{a} f \Delta L}$, where $f$ is the focal length of the slabs, $n_s$ is the effective refractive index of the slab mode, $\lambda_0$ is a center wavelength of the AWG, $n_\mathrm{a}$ is the group index of each waveguide of the waveguide array, and $\Delta L$ is the constant path-length difference between neighboring waveguides of the waveguide array \cite{okamoto2006}. In the SFWM process, photon pairs are generated that satisfy the frequency relationship $2 \nu_\mathrm{p} = \nu_\mathrm{s} + \nu_\mathrm{i}$, where $\nu_\mathrm{k} = c/\lambda_\mathrm{k}$ \textcolor{black}{($c$: the speed of light in a vacuum)}, and the phase matching condition \cite{matsuda16}. The pump, signal, and idler modes output from each input waveguide are focused at the output end of the second slab (arc o-o') with a spatial dispersion $\frac{\Delta x}{\Delta \lambda} = \frac{n_\mathrm{a} f \Delta L}{n_\mathrm{s} d \lambda_0}$. The dispersion of the focal spot is symmetric with respect to an exchange of the input and output port. We align the three photon pair sources ($\mathrm{I}_1 \sim \mathrm{I}_3$) to the three grid lines with a pitch $d$ on the arc i-i' around the center of the slab. Because of the dispersion in the AWG, the correlated photons created in each of the photon pair sources are focused onto different output positions. When the AWG is designed such that $m \Delta \nu = \left| \nu_\mathrm{p} - \nu_\mathrm{s} \right|$ ($m$: integer, $\Delta \nu$: AWG channel spacing in frequency), \textit{i.e.}, $m \Delta \lambda \simeq \lambda_0^2 \left| \frac{1}{\lambda_\mathrm{p}} - \frac{1}{\lambda_\mathrm{s}} \right|$ and $\lambda_0 = \lambda_\mathrm{p}$, the pump, signal and idler components are focused on the positions of the grids shown in Fig. 1(b), which shows the case of $m = 4$. Then, we connect output waveguides A$_j$ and B$_j$ ($j = 1\sim 3$) so as to collect the signal and idler photons.

In this configuration, the device serves as a compact high-dimensional path entanglement source on a chip. Since one WDM filter is shared by the photon-pair sources, fabrication errors similarly propagate to the spectral characteristics of the WDM channels, \textit{i.e.}, output photons. \textcolor{black}{A possible fabrication error of the AWG is a variation in the waveguide width, which modifies $\Delta \lambda$ via a variation in the group index $n_{\rm a}$. In our fabrication process described later in Sec. 3, we estimate $n_{\rm a}$ variation due to this fabrication error to be less than $10^{-3}$, which leads to $\Delta \lambda$ error to the same order.} Note that the arrangement of the i/o waveguides is not limited to the configuration shown in Fig. 1(b). \textcolor{black}{The largest single-stage AWG to date has 400 wavelength channels \cite{hibino02}, which indicates, in principle, that the dimension $N$ of more than 100 can be realized in our scheme.} Besides components shown in Fig. 1(a), bandpass filters can also be placed between the output waveguides and single photon detectors to eliminate unwanted frequency components of photons including the pump wavelength ($\lambda_\mathrm{p}$) components.

\section{Experimental setup}

We fabricated the essential part of the device and performed a proof-of-principle experiment of the scheme that generated entanglement for the minimum mode number $N = 2$.  \textcolor{black}{Correlated photons from neighboring waveguides have also been used for the generation of polarization entanglement \cite{sansoni17}.} We fabricated a photonic chip housing two silicon waveguides and an AWG with a waveguide core of low-nonlinear SiO$_x$ \textcolor{black}{($x \simeq$ 1.7 \cite{hiraki13})} using silicon-silica monolithic integration technology \cite{nishi10, matsuda14}. The chip, along with the experimental setup, is illustrated in Fig. 2(a).

Regarding the device fabrication, we fabricated silicon waveguides by electron-beam lithography and electron-cyclotron resonance (ECR) plasma etching on a silicon-on-insulator (SOI) substrate. The SiO$_{x}$ waveguides were fabricated in a region from which the top Si layer of the SOI substrate was removed by reactive ion etching (RIE). The SiO$_{x}$ layer in this region was deposited by low-temperature ECR plasma-enhanced chemical vapor deposition (PE-CVD). Next, the cores of the SiO$_{x}$ waveguides were fabricated by photolithography and RIE. Finally, the SiO$_{2}$ layer was deposited by ECR PE-CVD. The resulting SiO$_{x}$ waveguides had a core-cladding index contrast of $\sim$ 3\%. We also fabricated spot-size converters (SSCs) with a tapered Si waveguide for making a low-loss connection of the Si and SiO$_{x}$ waveguides. The SiO$_{x}$ waveguide was 3-\textmu m wide and 3-\textmu m thick. The silicon waveguides, having a rib structure 600-nm wide and 200-nm thick, respectively (see \cite{matsuda14}), are 1.37 cm long.

Pump pulses with a repetition rate of 100 MHz and pulse width of 200 ps are equally divided by an off-chip 3-dB fiber coupler, whose output ports are connected to an array of high-NA single mode fibers with a spacing of 127 \textmu m\textcolor{black}{, which corresponds to the separation of the input waveguides}. The output fields from the fiber array are coupled to the TE-polarized modes of the two silicon waveguides via SiO$_x$-based spot-size converters. \textcolor{black}{A variable attenuator is used to compensate for the imbalance of the device insertion losses between the mode 1 and 2.} The \textcolor{black}{input and output} coupling efficiency to the chip is estimated to be approximately $- 1$ dB/facet \textcolor{black}{\cite{matsuda14}}.

A correlated pair of signal and idler photons is created via the SFWM in the silicon waveguides with the pump pulses, and the pair is subsequently spectrally separated by the SiO$_x$ AWG on the end facet of the second slab. The design parameters of the AWG are as follows: $d =$ 30 \textmu m, $f =$ 1.75 mm and $\Delta L =$ 63 \textmu m; the total number of array waveguides is 100. With these parameter, the fabricated samples exhibited $\Delta \nu$ of 200 GHz and $\lambda_0$ = 1560.6 nm with a grating order of 53. The center wavelength of the pump pulses is set to $\lambda_\mathrm{p} = \lambda_0$. The two silicon waveguides (I$_1$ and I$_2$) are connected to the two central input ports of the AWG (Fig. 2(b)). Signal and idler photons are collected from ports A$_j$ and B$_j$ ($j = 1, 2$) with a channel separation $m$ of 3, and thus, a pump-to-signal (or -idler) detuning $m \Delta \nu$ of 600 GHz. \textcolor{black}{The insertion loss of the AWG is $-6.7$ dB including a $-0.35$ dB connection loss \cite{hiraki13} between Si and SiO$_x$ waveguides.} Figure 2(c) shows the TE-mode transmission spectra through the chip for the combinations of input and output waveguides used for the experiment. Good spectral overlaps in each transmission mode were obtained. The 3-dB passband width of the transmission window was approximately 90 GHz. \textcolor{black}{The pump leakage to the signal or idler modes is approximately $-$30 to $-$40 dB.}

\begin{figure*}[tb]
\centerline{\includegraphics[width=15cm]{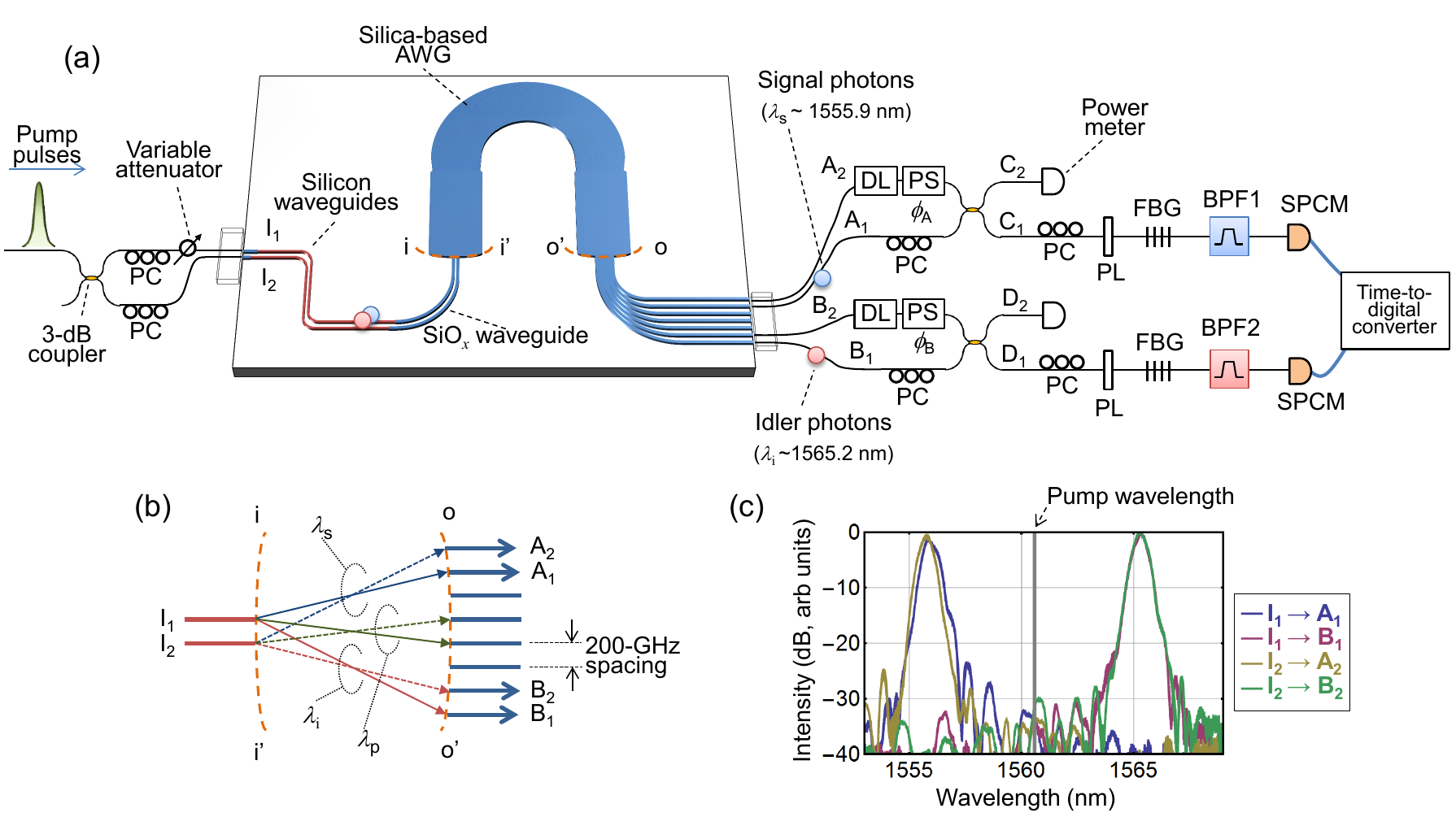}}
\caption{(a) Schematic diagram of the sample and the experimental setup. PC: polarization controller, DL: delay line, PS: phase shifter, PL: polarizer, FBG: fiber-Bragg grating, BPF: \textcolor{black}{tunable} band-pass filter, SPCM: single-photon counting module. The variable attenuator (b) I/O waveguide configurations of the device. (c) Transmission spectra of the AWG for TE polarization measured with an amplified spontaneous emission (ASE) source.}
\end{figure*}

The optical fields output from the chip are then collected by another high-NA fiber array. A fiber-pigtailed phase shifter and a 3-dB fiber coupler are used for the quantum state projection of the path-encoded states for each signal and idler mode. \textcolor{black}{Delay lines (DLs) are used to set the path-length difference in each interferometer to be zero}. After the unwanted wavelength components including pump wavelength are rejected with the fiber Bragg gratings (FBGs) and the band-pass filters (BPFs) (bandwidth: 0.9 nm (113 GHz)), photons output from one port of the 3-dB couplers are received by a single photon counting modules (SPCMs) (id210, ID Quantique SA) that operates at a gate frequency of 100 MHz, synchronized with the pump repetition rate. \textcolor{black}{The insertion losses of the spectral filters (an FBG and a BPF) and other components (including the 3 dB loss of a coupler) are $-$2.8 dB and $-$7 dB, respectively. With the spectral filters, we observe any photon counts from the leaked pump pulses. The total filter loss is mainly due to the loss of the tunable BPF and can be reduced to less than half by employing fixed low-loss filters.} The quantum efficiency, gate width, dark count rate, and dead time of the detectors are 21 $\%$, 1.0 ns, 2.1 kHz, and 10 $\mu$s, respectively. The coincidence rate is determined by measuring the time correlation of the signals output from the two SPCMs using a time-to-digital converter (id800). \textcolor{black}{The collection efficiency of the photons in each mode, which can be obtained as the total insertion loss of the fiber-optic components and the AWG, is $-$17.5 dB (excluding the detector efficiency).}

\section{Results and discussion}

To analyze the path-entangled state, we performed a coincidence measurement as a function of the relative phase differences $\phi_\mathrm{s}$ and $\phi_\mathrm{i}$ between the path-encoded states in signal and idler modes, respectively. To measure the phases, we retrieved the relative phases $\phi_\mathrm{A}$ and $\phi_\mathrm{B}$ of the pump frequency modes by observing the interference intensity of the pump pulses that leaked from the unused ports of the 3-dB couplers. As the wavelength difference between $\lambda_\mathrm{p}$ and $\lambda_\mathrm{s(i)}$ is less than 0.4 $\%$, we can assume that $\phi_\mathrm{A(B)} \simeq \phi_\mathrm{s(i)} + \Delta \phi_\mathrm{s(i)}$, where $\Delta \phi_\mathrm{s(i)}$ is the constant phase offset. The relative phases $\phi_\mathrm{A}$ and $\phi_\mathrm{B}$ naturally fluctuate because of the variation in the path lengths in the fiber-optic interferometers. Here, we utilized the phase fluctuation to automatically take data for various phase settings. We recorded the coincidence count $C$ while retrieving the relative phases every 200 ms and constructed the coincidence map $C\left( \phi_\mathrm{A}, \phi_\mathrm{B} \right)$.

The plots in Figure 3(a) show the normalized coincidence rate per second as a function of $\phi_\mathrm{A}$ and $\phi_\mathrm{B}$ over a measurement time of 24 hours. The phase bin size is $3^\circ$ but $45^\circ$ for the region where the pump interference intensity exhibits a small variation with respect to the change in phase. Data were discarded when the phase variation at the end of each measurement was larger than the phase bin size. We observed a clear two-photon interference fringe. In the ideal case, the coincidence rate is obtained as

\begin{equation}
C\left( \phi_\mathrm{A}, \phi_\mathrm{B} \right) = C_0 \left( 1 - V \cos(\phi_\mathrm{A} + \phi_\mathrm{B} + \Delta \phi) \right),
\end{equation}
\noindent
where $\Delta \phi$ is the constant offset associated with the wavelength difference between the pump and signal (or idler) modes, $V$ is the fringe visibility and $C_0$ is a coefficient. Using this function, we performed a fitting of the experimental data, in which errors in the count rate (Poisson statistics assumed) and each phase (bin sizes \textcolor{black}{including the large ($45^\circ$) bins)} were taken into account \textcolor{black}{(and the same hereinafter)}. \textcolor{black}{From the fitting, we obtained the curved surface shown in Fig. 3(a) with} $V = 0.768 \pm 0.002$ (and with the accidentals subtraction $V = 0.813 \pm 0.002$). Thus, we successfully obtained the entangled state with the visibility value larger than that associated with the violation of Bell's inequality ($1/\sqrt{2}$).

Figure 3(b) plots slices of the coincidence rate at $\phi_\mathrm{A} = 51^\circ$ and $141^\circ$, which are non-orthogonal to each other. Both slices show clear interference fringes with visibilities of $V = 0.762 \pm 0.018$ and $V = 0.771 \pm 0.014$, respectively. \textcolor{black}{Next, we extract $\left| S \right|$ values for the Clauser-Horne-Shimony-Holt (CHSH) Bell inequality \cite{chsh} from the coincidence map in Fig. 3(a) for all the possible phase combinations. We find the maximum $\left| S \right|$ value to be $2.26 \pm 0.14$ (accidentals included), which demonstrate that the generated state can violate Bell's inequality. The small violation is due to the small coincidence counts per phase bin.} 

\begin{figure}[bt]
\centerline{\includegraphics[width=15cm]{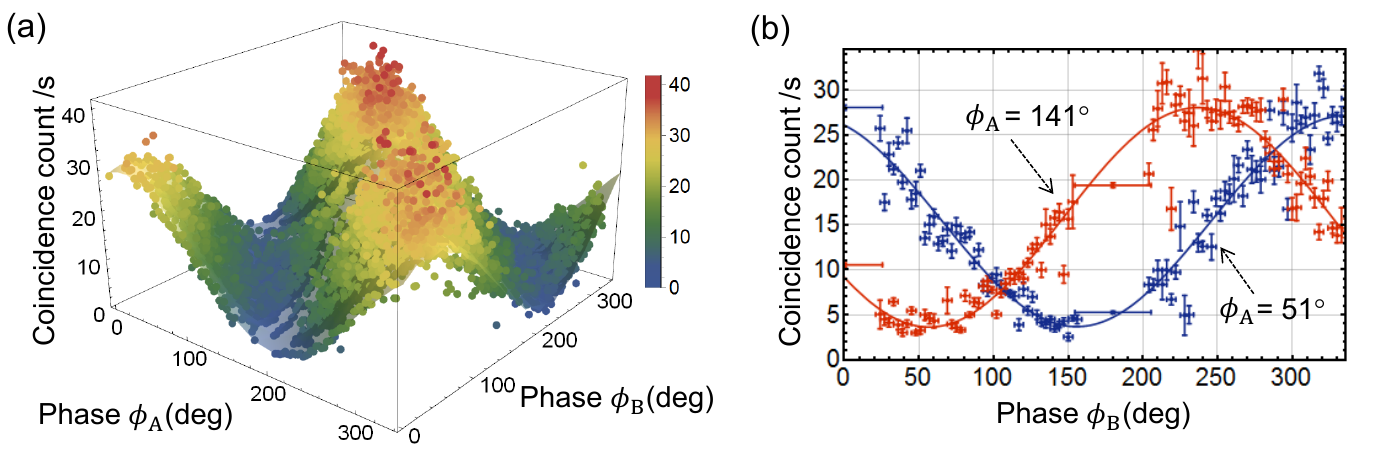}}
\caption{(a) Measured normalized coincidence count as a function of relative phases of signal and idler modes retrieved from the interferometric signal. (b) Slices of the coincidence map in the two non-orthogonal measurement basis.}
\end{figure}
 
Now let us investigate the effect of the subtle mismatch of AWG transmission spectra (Fig. 2(c)) on the imperfection in the visibility values. Using the joint spectral amplitude of photon pairs $S_j ( \omega_\mathrm{s}, \omega_\mathrm{i} )$, the two-photon state at the output of the chip can be written as

\begin{equation}
	\ket{\psi} = \frac{1}{\sqrt{2}} \sum_{j = 1}^2 \int \int d\omega_\mathrm{s} d\omega_\mathrm{i} S_j \left( \omega_\mathrm{s}, \omega_\mathrm{i} \right) \hat{a}_j^\dagger (\omega_\mathrm{s}) \hat{b}_j^\dagger (\omega_\mathrm{i}) \ket{0},
\end{equation}

\noindent
\textcolor{black}{where $\omega_{\rm j} = 2 \pi \nu_{\rm j}$.} Measurement basis of the photon pairs at the fiber modes C$_1$ and D$_1$ in Fig. 2(a) is $\ket{\xi} = \hat{c}_1^\dagger (\omega) \hat{d}_1^\dagger (\omega') \ket{0}$, where creation operator $\hat{c}_1^\dagger$ in fiber mode C$_1$ is $\hat{c}_1^\dagger = \frac{1}{\sqrt{2}}\left( \mathrm{e}^{- i \phi_s} \hat{a}_1^\dagger + i \hat{a}_2^\dagger \right)$, and similarly for $\hat{d}_1^\dagger$ in fiber mode D$_1$. The integral in Eq. (3) can be taken from $- \infty $ to $+ \infty $ since the bandwidths of the BPFs are larger than the passband width of the AWG. \textcolor{black}{Then we obtain the coincidence probability as $P_c = \int \int d\omega d\omega' \left|  \braket{\xi}{\psi} \right|^2 = \int \int d\omega d\omega' \left| \frac{\mathrm{e}^{i (\phi_{\rm s} + \phi_{\rm i}) }}{2} S_1 ( \omega, \omega' ) + \frac{1}{2} S_2 ( \omega, \omega' ) \right|^2$.}

Because the pump bandwidth (2.2 GHz assuming a transform-limited Gaussian pulse) is much narrower than the collection bandwidth of signal and idler photons ($\sim$ 90 GHz),  here we assumed that pump beam to be quasi-cw. \textcolor{black}{We show the shape of the joint spectral intensity $ \left| S_1 ( \omega_s, \omega_i ) \right|^2$ as an example in Fig. 4.} Accordingly, the joint spectral amplitude $S_j ( \omega_s, \omega_i ) = f_j (\omega_s) g_j (\omega_i) \delta ( 2\omega_p - \omega_s - \omega_i) $, where $\omega_p$ denotes the center angular frequency of the pump and $f_j (\omega)$ and $g_j (\omega)$ are the transmission spectra of the optical field output into modes A$_j$ and B$_j$. Substituting this into $P_c$ and comparing with Eq. (2), we find 

\begin{equation}
	V = \frac{2 \int d\omega \mathrm{Re} \left[ f_1 ( \omega ) g_1 ( 2\omega_p - \omega ) f_2 ( \omega ) g_2 ( 2\omega_p - \omega ) \right]}{\int d\omega \left( \left| f_1 ( \omega ) g_1 ( 2\omega_p - \omega ) \right|^2 + \left| f_2 ( \omega ) g_2 ( 2\omega_p - \omega ) \right|^2 \right)},
\end{equation}

\noindent
Using the square roots of the transmission spectra shown in Fig. 2(c) for $f_j (\omega)$ and $g_j (\omega)$ (assuming no phase dispersion), we obtain $V = 96.8\% $. This indicates that our multi-input multi-output AWG has good spectral overlaps for generating high-degree of entanglement. Other possible reasons for the imperfection in the visibilities are the polarization drift in fiber optics during the measurement and high frequency component of the phase variation with an amplitude larger than the phase bin sizes. \textcolor{black}{Note that the experimental system was not temperature controlled.} These issues will be eliminated by integrating the interferometer for the state projection on the same chip with SiO$_x$ waveguides.

\begin{figure}[bt]
\centerline{\includegraphics[width=8cm]{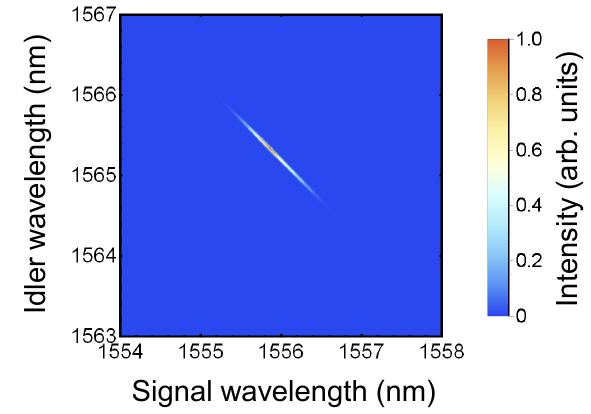}}
\caption{\textcolor{black}{Joint spectral intensity $\left| S_1 \left(\omega_{\rm s}, \omega_{\rm_i} \right) \right|^2$ of photons generated in mode 1 calculated with the AWG transmission spectra shown in Fig. 2(c) and the spectral width of the pump pulses.}}
\end{figure}

\section{Conclusion}
We proposed a scheme to generate high-dimensional path-encoded entangled states of photons by using an AWG with nonlinear waveguide inputs and demonstrated it on a Si-silica monolithic integration platform. To further extend the dimensions of entanglement to be characterized, it is necessary to integrate $N \times N$ reconfigurable unitary circuits on the same chip for the phase stabilization. The output waveguides of the silica-based AWG can be directly used as interfaces with such circuits. By incorporating fabrication technologies for large \cite{hibino02} and low-loss \cite{sugita00} AWGs, our scheme can provide a compact source for large-scale high-dimensional photonic entanglement on a chip.

\section*{Acknowledgement}
NM is grateful to Manabu Oguma for fruitful discussions. This work was supported by JSPS KAKENHI Grant Number JP26706021.


\section*{References}

\begin{thebibliography}{99}

\bibitem{nielsen00} M. A. Nielsen, I. L. Chuang, ``Quantum computation and quantum information,'' Cambridge University Press, Cambridge (2000).


\bibitem{cerf02} N. J. Cerf, M. Bourennane, A. Karlsson, and N. Gisin, ``Security of Quantum Key Distribution Using $\mathit{d}$-Level Systems,'' \prl {\bfseries 88}, 127902 (2002).

\bibitem{xu16} F. Xu, J. H. Shapiro, and F. N. C. Wong, ``Experimental fast quantum random number generation using high-dimensional entanglement with entropy monitoring,'' Optica {\bf 3}, 1266-1269 (2016).

\bibitem{lanyon09} B. P. Lanyon, M. Barbieri, M. P. Almeida, T. Jennewein, T. C. Ralph, K. J. Resch, G. J. Pryde, J. L. O{\textquoteright}Brien, A. Gilchrist, A. G. White, ``Simplifying quantum logic using higher-dimensional Hilbert spaces,'' Nat. Phys. {\bf 5}, 134--140 (2009).

\bibitem{hu16} X.-M. Hu, J.-S. Chen, B.-H. Liu, Y. Guo, Y.-F. Huang, Z.-Q. Zhou, Y.-J. Han, C.-F. Li, and G.-C. Guo, ``Experimental test of compatibility-loophole-free contextuality with spatially separated entangled qutrits,'' \prl {\bfseries 117}, 170403 (2016).

\bibitem{dada11} A. C. Dada, J. Leach, G. S. Buller, M. J. Padgett, and E. Andersson, ``Experimental high-dimensional two-photon entanglement and violations of generalized Bell inequalities,'' \natphy {\bfseries 7}, 677--680 (2011).

\bibitem{krenn14} M. Krenn, M. Huber, R. Fickler, R. Lapkiewicz, S. Ramelow, and A. Zeilinger, ``Generation and confirmation of a (100 $\times$ 100)-dimensional entangled quantum system,'' Proc. Natl. Acad. Sci. {\bfseries 111}, 6243--6247 (2014).

\bibitem{zhang16} Y. Zhang, F. S. Roux, T. Konrad, M. Agnew, J. Leach, and A. Forbes, ``Engineering two-photon high-dimensional states through quantum interference,'' \sciadv {\bfseries 2}, e1501165 (2016).

\bibitem{riedmatten02} H. de Riedmatten, I. Marcikic, H. Zbinden, and N. Gisin, ``Creating high dimensional time-bin entanglement using mode-locked lasers,'' Quant. Inf. Comp. {\bfseries 2}, 425--433 (2002).

\bibitem{ikuta16} T. Ikuta and H. Takesue, ``Enhanced violation of the Collins-Gisin-Linden-Massar-Popescu inequality with optimized time-bin-entangled ququarts,'' Phys. Rev. A {\bfseries 93}, 022307 (2016).

\bibitem{xie15} Z. Xie, T. Zhong, S. Shrestha, X. Xu, J. Liang, Y.-X. Gong, J. C. Bienfang, A. Restelli, J. H. Shapiro, F. N. C. Wong, and C. W. Wong, ``Harnessing high-dimensional hyperentanglement through a biphoton frequency comb,'' \natpho {\bfseries 9}, 536 (2015).

\bibitem{kues17} M. Kues, C. Reimer, P. Roztocki, L. R. Cort{\'e}s, S. Sciara, B. Wetzel, Y. Zhang, A. Cino, S. T. Chu, B. E. Little, D. J. Moss, L. Caspani, J. Aza\~{n}a, R. Morandotti, ``On-chip generation of high-dimensional entangled quantum states and their coherent control,'' Nature {\bf 546}, 622--626 (2017).

\bibitem{imany17} P. Imany, J. A. Jaramillo-Villegas, O. D. Odele, K. Han, D. E. Leaird, M. Qi, A. M. Weiner, ``High-dimensional frequency-bin entangled photons in an optical microresonator on a chip,'' arXiv:1707.02276 [quant-ph].

\bibitem{schaeff12} C. Schaeff, R. Polster, R. Lapkiewicz, R. Fickler, S. Ramelow, and A. Zeilinger, ``Scalable fiber integrated source for higher-dimensional path-entangled photonic quNits,'' \opex {\bfseries 20}, 16145--16153 (2012).

\bibitem{schaeff15} C. Schaeff, R. Polster, M. Huber, S. Ramelow, and A. Zeilinger, ``Experimental access to higher-dimensional entangled quantum systems using integrated optics,'' Optica {\bfseries 2}, 523--529 (2015).

\bibitem{politi08} A. Politi, M. J. Cryan, J. G. Rarity, S. Yu, J. L. O{\textquoteright}Brien, ``Silica-on-silicon waveguide quantum circuits,'' Science {\bfseries 5876}, 1500--1503 (2008).

\bibitem{peruzzo10} A. Peruzzo, M. Lobino, J. C. F. Matthews, N. Matsuda, A. Politi, K. Poulios, X.-Q. Zhou, Y. Lahini, N. Ismail, K. W$\rm \ddot{o}$rhoff, Y. Bromberg, Y. Silberberg, M. G. Thompson, and J. L. O'Brien, ``Quantum walks of correlated photons,'' \sci {\bf 329}(5998), 1500--1503 (2010).

\bibitem{spring13} J. B. Spring, B. J. Metcalf, P. C. Humphreys, W. S. Kolthammer, X.-M. Jin, M. Barbieri, A. Datta, N. Thomas-Peter, N. K. Langford, D. Kundys, J. C. Gates, B. J. Smith, P. G. R. Smith, and I. A. Walmsley, ``Boson sampling on a photonic chip,'' \sci {\bf 339}(6121), 798--801 (2013).

\bibitem{spagnolo14} N. Spagnolo, C. Vitelli,	M. Bentivegna, D. J. Brod, A. Crespi, F. Flamini, S. Giacomini, G. Milani, R. Ramponi, P. Mataloni, R. Osellame, E. F. Galv\~ao, and F. Sciarrino, ``Experimental validation of photonic boson sampling,'' Nat. Photon. {\bf 8}, 615--620 (2014).

\bibitem{carolan15} J. Carolan, C. Harrold, C. Sparrow, E. Mart{\'\i}n-L{\'o}pez, N. J. Russell, J. W. Silverstone, P. J. Shadbolt, N. Matsuda, M. Oguma, M. Itoh, G. D. Marshall, M. G Thompson, J. C. F. Matthews, T. Hashimoto, J. L. O{\textquoteright}Brien, A. Laing, ``Universal linear optics,"Science {\bfseries 349}, 711--716 (2015).

\bibitem{reck94} M. Reck, A. Zeilinger, H. J. Bernstein, and P. Bertani, ``Experimental realization of any discrete unitary operator,'' \prl {\bfseries 73}, 58 (1994).

\bibitem{harris16} N. C. Harris, D. Bunandar, M. Pant, G. R. Steinbrecher, J. Mower, M. Prabhu, T. Baehr-Jones, M. Hochberg, and D. Englund, ``Large-scale quantum photonic circuits in silicon,'' Nanophotonics {\bfseries 5}, 3 (2016).

\bibitem{ding16} Y. Ding, D. Bacco, K. Dalgaard, X. Cai, X. Zhou, K. Rottwitt, and L. K. Oxenl{\o}we, ``High-dimensional quantum key distribution based on multicore fiber using silicon photonic integrated circuits,'' npj Quantum Information {\bf 3}, 25 (2017).

\bibitem{canas16} G. Ca{\~n}as, N. Vera, J. Cari{\~n}e, P. Gonz{\'a}lez, J. Cardenas, P. W. R. Connolly, A. Przysiezna, E. S. G{\'o}mez, M. Figueroa, G. Vallone, P. Villoresi, T. Ferreira da Silva, G. B. Xavier, and G. Lima, ``High-dimensional decoy-state quantum key distribution over 0.3 km of multicore telecommunication optical fibers,'' arXiv:1610.01682 [quant-ph].


\bibitem{matsuda12} N. Matsuda, H. Le Jeannic, H. Fukuda, T. Tsuchizawa, W. J. Munro, K. Shimizu, K. Yamada, Y. Tokura, and H. Takesue, ``A monolithically integrated polarization entangled photon pair source on a silicon chip,'' \scirep {\bf 2}, 817 (2012).

\bibitem{kaiser12} F. Kaiser, A. Issautier, L. A. Ngah, O. Danila, H. Herrmann, W. Sohler, A. Martin, and S. Tanzilli, ``High-quality polarization entanglement state preparation and manipulation in standard telecommunication channels,'' \njp {\bf 14}, 085015 (2012).

\bibitem{autebert16} C. Autebert, J. Trapateau, A. Orieux, A. Lemaitre., C. Gomez-Carbonell, E. Diamanti, I. Zaquine, and S. Ducci, ``Multi-user quantum key distribution with entangled photons from an AlGaAs chip,'' Quantum Sci. Technol. {\bf 1}, 01LT02 (2016).

\bibitem{suzuki98} S. Suzuki, A. Himeno, M. Ishii, ``Integrated multichannel optical wavelength selective switches incorporating an arrayed-waveguide grating multiplexer and thermooptic switches,'' J. Lightwave Technol. {\bf 16}, 650--955 (1998).


\bibitem{nishi10} H. Nishi, T. Tsuchizawa, T. Watanabe, H. Shinojima, S. Park, R. Kou, K. Yamada, and S. Itabashi, ``Monolithic integration of a silica-based arrayed waveguide grating filter and silicon variable optical attenuators based on p-i-n carrier-injection structure,'' Appl. Phys. Express {\bf 3}, 102203 (2010).

\bibitem{sharping06} J. E. Sharping, K. F. Lee, M. A. Foster, A. C. Turner, B. S. Schmidt, M. Lipson, A. L. Gaeta, and P. Kumar, ``Generation of correlated photons in nanoscale silicon waveguides,'' \opex {\bf 14}(25), 12388--12393 (2006).

\bibitem{takesue07} H. Takesue, Y. Tokura, H. Fukuda, T. Tsuchizawa, T. Watanabe, K. Yamada, and S. Itabashi, ``Entanglement generation using silicon wire waveguide,'' \apl {\bf 91}(20), 201108 (2007).

\bibitem{silverstone14} J. W. Silverstone, D. Bonneau, K. Ohira, N. Suzuki, H. Yoshida, N. Iizuka, M. Ezaki, C. M. Natarajan, M. G. Tanner, R. H. Hadfield, V. Zwiller, G. D. Marshall, J. G. Rarity, J. L. O{\textquoteright}Brien, M. G. Thompson, ``On-chip quantum interference between silicon photon-pair sources,'' Nat. Photon. {\bf 8}, 104--108 (2014).

\bibitem{matsuda14} N. Matsuda, P. Karkus, H. Nishi, T. Tsuchizawa, W. J. Munro, H. Takesue, K. Yamada, "On-chip generation and demultiplexing of quantum correlated photons using a silicon-silica monolithic photonic integration platform," Opt. Express {\bf 22}, 22831 (2014).

\bibitem{matsuda16} N. Matsuda, H. Takesue, "Generation and manipulation of entangled photons on silicon chips," Nanophotonics {\bf 5}, 440 (2016). 

\bibitem{okamoto2006} K. Okamoto, ``Fundamentals of optical waveguides, 2nd edition,'' Academic Press (2010).

\bibitem{hibino02} Y. Hibino, ``Recent advances in high-density and large-scale AWG multi/demultiplexwers with higher index-constrast silica-based PLCs,'' IEEE J. Selected Topic Quantum Electron {\bf 8}, 1090--1101 (2002).

\bibitem{sansoni17} L. Sansoni, K. H. Luo, C. Eigner, R. Ricken, V. Quiring, H. Herrmann, C. Silberhorn, ``A two-channel, spectrally degenerate polarization entangled source on chip,'' npj Quantum Information {\bf 3}, 5 (2017).

\bibitem{hiraki13} T. Hiraki, H. Nishi, T. Tsuchizawa, R. Kou, H. Fukuda, K. Takeda, Y. Ishikawa, K. Wada, K. Yamada, ``Si-Ge-Silica Monolithic Integration Platform and Its Application to a 22-Gb/s $\times$ 16-ch WDM Receiver,'' IEEE Photonics Journal {\bf 5}, 4500407 (2013).

\bibitem{chsh} J. F. Clauser, M. A. Horne, A. Shimony, R. A. Holt, ``Proposed experiment to test local hidden-variable theories,'' Phys. Rev. Lett. {\bf 23}, 880 (1969).
\bibitem{sugita00} A. Sugita, A. Kaneko, K. Okamoto, M. Itoh, A. Himeno, and Y. Ohmori, ``Very low insertion loss arrayed-waveguide grating with vertically tapered waveguides,'' IEEE Photon. Technol. Lett. {\bf 12}, 1180 (2000).


\end{thebibliography}
\end{document}